
\input psbox.tex
\splitfile{\jobname}
\let\autojoin=\relax
\magnification=1200
\hfuzz=1pt
\input psbox.tex

\font\bigbf=cmbx12 at 14truept
\font\msxm=msxm10

\baselineskip = 12pt

\def\firstskip{\vskip20truept}
\def\onelnskp{\vskip10truept}
\def\br{{\bf r}}
\def\rint{{\rm int}}
\def\got{\hbox{\msxm \char38}}
\def\lot{\hbox{\msxm \char46}}
\def\half{\hbox{$\textstyle {1\over 2}$}}

\def\etal{{\it et al.}}
\def\la{\langle}
\def\ra{\rangle}
\def\deg{$^\circ$}
\def\tel{\tilde\epsilon_1}
\def\ep{\epsilon}
\def\eps{\epsilon}

\def\0{{\vec 0}}
\def\A{{\vec A}}

\def\j{{\vec j}}
\def\f{{\vec f}}
\def\r{{\vec r}}
\def\R{{\vec R}}

\def\vv{{\vec v}}

\def\CD{{\cal D}}

\def\CH{{\cal H}}

\def\CO{{\cal O}}
\def\CP{{\cal P}}
\def\CZ{{\cal Z}}

\noindent
Lectures presented at the NATO Advanced Study Institute: ``Phase Transitions
and Relaxation in Systems with Competing Energy Scales,'' April 13--23,
1993, Geilo, Norway.
\vskip 20truept
\noindent
{\bigbf Vortex Line Fluctuations in Superconductors from
Elementary Quantum Mechanics}

\vskip39truept

\noindent
\hskip5truepc\vbox{\parindent=0pt\hsize=30truepc
David R.~Nelson\hfill\break
{\it Lyman Laboratory of Physics}\hfill\break
{\it Harvard University}\hfill\break
{\it Cambridge, Massachusetts \ 02138}}

\vskip33truept

\noindent
ABSTRACT. Concepts from elementary quantum mechanics can be used to
understand vortex line fluctuations in high-temperature superconductors.
Flux lines are essentially classical objects, described by a string
tension, their mutual repulsion, and interactions with pinning centers.
The classical partition function, however, is isomorphic to the imaginary
time path integral description of quantum mechanics. This observation is
used to determine the thermal renormalization of critical currents,
the decoupling field,
the flux lattice melting temperature at low and moderate inductions,
and to estimate the degree of entanglement in dense flux liquids. The
consequences of the ``polymer glass'' freezing scenario, which assumes that
the kinetic constraints of entanglement prevent field cooled flux liquids
from crystallizing, are reviewed.

\firstskip

\noindent
{\bf 1. Introduction}

\onelnskp

\noindent
The past five years have been a time of great ferment in the study of
high-temperature superconductors. Although no general consensus about the
microscopic mechanism for superconductivity in the cuprate materials has
emerged, there is now considerable understanding of the remarkable behavior of
these materials in a magnetic field. The theoretical analysis of vortex
fluctuations in these extreme Type II materials requires only an
underlying Ginzburg-Landau theory with a BCS-like order parameter.
The basic conclusions are {\it independent} of the precise microscopic
mechanism, and rely instead on the remarkably different Ginzburg-Landau
parameters (coherence length, temperature range and anisotropy) which
distinguish the cuprates from their low $T_c$ counterparts.

\topinsert
\centerline{\psboxto(0pt;5truein){fig1.eps}}
\onelnskp
\noindent
Figure 1. Schematic phase diagram of a high temperature superconductor in the
clean limit. Although quantized vortex lines appear below $B^0_{c2}(T)$, the
Abrikosov crystal phase appears only below $T_m(B)$. The Meissner phase
collapses to the heavy line along the temperature axis in this representation.
Point disorder only affects the flux liquid below the dotted line. Ref.~[8]
presents arguments that the maximum in the melting curve actually bends
down closer to $T_c$ than indicated here.
\onelnskp
\endinsert

Figure 1 shows a schematic temperature-magnetic field phase diagram for
cuprate superconductors subjected only to weak point disorder in the form
of oxygen vacancies. Throughout this paper we assume for simplicity a
field oriented along the $c$ axis, perpendicular to the copper-oxide
planes. The magnetic field $B$ (proportional to the vortex density) is
plotted because demagnetizing corrections in the usual slab-like
experimental geometry insure that $B$ rather than the magnetic induction
$H$ is held fixed in most experimental situations. The famous
Abrikosov flux lattice, which exists for all temperatures below the
upper critical field $B_{c2}(T)$ in  mean field theory [1], appears
here only below a much lower ``melting line'' $T_m(B)$. Above this line,
melted vortex arrays entangle in a novel ``flux liquid.''
The line $B^0_{c2}(T)$
marks the onset of enhanced diamagnetism but is not expected to be a sharp
phase boundary. The crossover fields $B_{c1}$ and $B_x$ are discussed in
Sec.~3.  There is now considerable evidence [2--5]
that the Abrikosov lattice in clean (twin-free) single crystals of
yittrium barium copper oxide (YBCO) melts at $T_m(B)$ via a
first-order phase transition [6] into a flux liquid in which the
quantized vortex filaments presumably
wander and entangle in a complicated fashion.
Weak point disorder significantly alters the properties of the flux liquid
only below the dotted line [7]. Note that a sliver of flux liquid may
exist above this line and below the melting curve down to quite low
temperatures. It is still unclear whether point disorder alone is sufficient
to produce a distinct thermodynamic ``vortex glass'' phase~[8] below
the dotted line or within the crystalline region.

\topinsert
\centerline{\psboxto(4truein;0pt){fig2.eps}}
\onelnskp
\noindent
Figure 2. Effect of columnar pins on the irreversibility line of
Tl$_2$Ba$_2$Ca$_2$Cu$_3$O$_{10}$ [9]. In the absence of irradiation,
the resistivity only becomes unmeasurably small in the shaded region
below $T_{ir}(B)$. The reentrant low field vortex liquid regime of
Fig.~1 appears for $B\lot10^{-2}$~T, and is hence not visible on this
scale. Figure courtesy of R. Budhani, Brookhaven National Laboratory.
\onelnskp
\endinsert

Figure 2 shows another striking development---a remarkable shift in the
``irreversibility line'' of a highly anisotropic thallium-based compound
engineered via the deliberate introduction of columnar pins [9]. Similar
pinning centers have been injected by many groups via heavy ion irradiation
[10--13] with sufficient energy to produce long tracks of damaged material.
The ``irreversibility line'' $T_{ir}(B)$ is the boundary below which the
dynamics of field cooled materials slows down drastically [14]. In samples
{\it without} correlated disorder the melting curve and irreversibility
line may in fact be almost identical [8].

Since $T_{ir}(B)$ coincides approximately with the temperature at which
the resistivity becomes unmeasurably small, the large upward shift in
Fig.~2 is of considerable technological as well as intellectual interest.
Note for example that the effective critical temperature for superconductivity
at $B=2.7$ tesla shifts from 53~K to 87~K upon irradiation. The critical
current at liquid nitrogen temperature (77~K) increases by three orders
of magnitude [9]. It is believed that after irradiation the irreversibility
line becomes a locus of thermodynamically sharp ``bose glass'' transitions,
below which the flux lines are localized on the columnar defects [15]. Other
forms of correlated disorder may be important even in unirradiated samples.
Twin boundaries, for example, appear to be responsible for the apparently
continuous transition observed in twinned YBCO samples with the field
aligned parallel to the $c$ axis [3].

In these lectures, we highlight these developments by describing how flux lines
interact with correlated pinning centers and with each other. To account
for thermal fluctuations, one must average over vortex configurations.
Because the configuration sums bear a strong resemblance to the imaginary time
path
integral formulation of quantum mechanics [16], wave functions and binding
energies
which appear in simple quantum problems (such as the square well and
the harmonic oscillator) have important implications for flux lines.
With this identification we can quickly compute the free energy and
localization length of vortices in the bose glass phase, the corresponding
critical currents, and determine as well the decoupling field and
line of melting temperatures in
clean systems [15]. In the flux liquid, we can readily estimate the degree
of entanglement of the wandering vortex lines. If entanglement occurs
in temperature and field ranges where flux cutting barriers are high, it
may prevent crystallization and lead to a polymer-like glass transition
even in the absence of strong pinning disorder [17]. The experimental
consequences of this scenario are reviewed at the end of this brief,
elementary review.

The idea of using Schroedinger's equation to solve problems in
classical statistical mechanics is not new. It was used by S.F. Edwards,
P.G. deGennes and others starting in the 1960s to treat the conformations of
flexible polymer chains [18]. One can go further with this analogy for flux
lines, however, because these are equivalent to a system of {\it directed}
polymers, with a common average orientation. As a result,
distant self-interactions along the filaments
can usually be neglected, in contrast to isotropic self-avoiding
polymer solutions. The essential physics is captured if one introduces a line
tension to control vortex wandering and allows interactions between
{\it different} polymers as well as interactions with the relevant pinning
centers. Although we focus here primarily on simple one and two-line
problems, many flux line statistical mechanics (with multiple pins) is in fact
equivalent to the many body quantum mechanics of bosons in two
dimensions [15, 17]. This system differs from the otherwise closely related
problem of helium films on disordered substrates [19] because vortices behave
dynamically like {\it charged} bosons, and are easily manipulated in
experiments by the
injection of supercurrents. A simple explanation of why vortex
probability distributions are boson-like in a thick sample can
be found in Sec.~2.3. The richness and complexity of the {\it many} flux line
problem rivals the physics of correlated electrons in semiconductors and
metals. For details, readers are referred to Refs.~15 and 17, and a recent
review [20].

\firstskip

\noindent{\bf 2. Correlated Pinning and Quantum Bound States}

\onelnskp

\noindent
2.1. MODEL-FREE ENERGY

\onelnskp

\noindent
We start with a model-free energy $F_N$ for $N$ flux lines in a sample of
thickness
$L$, defined by their trajectories $\{\r_j(z)\}$ as they traverse a sample with
both columnar pins and external magnetic field aligned with the $z$-axis, i.e.,
in the
direction perpendicular to the CuO$_2$ planes,

\eject
$$
\eqalignno{
F_N=&{1\over 2}\tilde\ep_1\sum_{j=1}^N\int_0^L
\left|{d\r_j(z)\over dz}\right|^2\;dz+
{1\over 2}\sum_{i\not=j}\int_0^L
V(|\r_i(z)-\r_j(z)|)\;dz\cr
&+\sum_{j=1}^N\int_0^LV_D[\r_j(z)]\;dz&(2.1{\rm a})\cr}
$$
with
$$
V_D(\r)=\sum_{k=1}^M \,V_1(\r-\R_k)\;.
\eqno(2.1{\rm b})
$$
Here
$V(|\r_i-\r_j|)=2\ep_0\,K_0(r/\lambda_{ab})$, is the interaction
potential between lines with in-plane London penetration depth $\lambda_{ab}$
and the random potential $V_D(\r)$ arises from a $z$-independent set of $M$
disorder-induced columnar pinning potentials $V_1(\r)$ centered on sites
$\{\R_k\}$. The tilt modulus $\tilde\ep_1\approx
(M_\perp/M_z)\ep_0\ln(\lambda_{ab}
/\xi_{ab})$,
where the material anisotropy is embodied in the effective mass ratio
$M_\perp/M_z\ll 1$, and $\ep_0\approx(\phi_0/4\pi\lambda_{ab})^2$ is the energy
scale for the interactions.
The potential $V_D(\r)$ arises from  identical cylindrical traps  assumed
for simplicity to pass
completely through the sample with well depth per unit
length $U_0$ and effective radius $b_0$. The parameter
$
b_0\approx\max\{c_0,\xi_{ab}\}
$
where
$c_0\approx 25-40$~\AA\ is the radius of the columnar pins and $\xi_{ab}$ is
the
superconducting coherence length in the $ab$-plane.

A complete analysis of the many-line statistical mechanics associated with
Eqs.~(2.1) requires multiple path integrals over vortex trajectories weighted
by
$e^{-F_N/T}$ and subject to a complicated random pinning potential. See
Fig.~3.  Our goal here is to illuminate the essential physics by studying a few
simple problems involving one or two flux lines and only a few columnar pins.

\topinsert
\centerline{\psboxto(\hsize;0pt){fig3.eps}}
\onelnskp
\noindent
Figure 3. Schematic of columnar pins and pinned vortex lines.
\onelnskp
\endinsert

\onelnskp

\noindent
2.2. ONE FLUX LINE AND ONE COLUMNAR PIN

\onelnskp

\noindent
Consider a vortex line trapped near a single columnar pin
with well depth $U_0$ and radius $b_0$ parallel to $z$ in an
otherwise defect-free sample of thickness $L$, as shown in Fig.~4. A quantity
of considerable physical interest is
the binding {\it free} energy per unit length
$$
U(T)=U_0-TS\;,
\eqno(2.2)
$$
where $S$ is the entropy reduction due to confinement. This free energy is
given
by a path integral,
$$
e^{U(T)L/T}=
{
\int\CD \r(z)\exp\left[
-{\tilde \ep_1\over 2T}\int_0^L\left({d\r\over dz}\right)^2dz
-{1\over T}\int_0^L V_1[\r(z)]\;dz\right]\over
\int\CD r(z)\exp\left[-
{\tilde \ep_1\over 2T}\int_0^L
\left({d\r\over dz}\right)^2\;dz
\right]}\;,
\eqno(2.3)
$$
where the denominator is required to subtract off the entropy of an unconfined
line far from the pin. The cylindrically symmetric confining potential
$V_1(r)$, indicated in Fig.~4b, tends to zero as $|\r|\rightarrow \infty$.
The path integrals in (2.3) follow from standard statistical methods which
express them in terms of the eigenvalues of a transfer matrix. In the limit
$L\rightarrow\infty$, the smallest eigenvalue dominates, and $U(T)=-E_0(T)$,
where $E_0(T)$ is the ground state energy of a two-dimensional ``Schroedinger
equation'' (see Appendix~A),
$$
\left[-{T^2\over 2\tilde \ep_1}
\nabla_\perp^2+V_1(r)\right]
\psi_0(\r)=
E_0\psi_0(\r)\;.
\eqno(2.4)
$$
Here and henceforth, all vectors $\r$ will refer to positions in the plane
perpendicular to $\hat z$ Note that $T$ plays the role of the Planck parameter
$\hbar$ and $\tilde\ep_1$ plays the role of mass $m$ in this quantum mechanical
analogy.

\topinsert
\centerline{\psboxto(\hsize;0pt){fig4.eps}}
\onelnskp
\noindent
Figure 4. One columnar pin and one vortex line, indicating (a)~the localization
length $\ell_\perp(T)$ and (b)~the thermally renormalized pinning potential
$U(T)$.
\onelnskp
\endinsert

The ground state wave function $\psi_0(\r)$ determines the localization length
$\ell_\perp(T)$ displayed in Fig.~3a. As shown in Appendix~B, the probability
$P(\r)$ of finding a point on the vortex
at transverse displacement $\r$ relative to the center of the pin is
independent
of $z$ and given by the square of $\psi_0(\r)$, just as in elementary quantum
mechanics,
$$
\CP(\r)=\psi_0^2(\r)\Bigg/\int d^2r\psi_0^2(\r)\;.
\eqno(2.5)
$$
Because (2.4) is unchanged under complex conjugation, $\psi_0(\r)$ can
always be chosen to be real. We then define the localization length as
$$
\ell_\perp^2=\int d^2r\;r^2\psi_0^2(r)\Bigg/
\int d^2r\;\psi_0^2(r)\;.
\eqno(2.6)
$$

The properties of a vortex near a columnar pin now follow from standard results
for a
quantum particle in a cylindrical potential [21].
If we assume for simplicity a cylindrical square well, ($V_1(r)\equiv -U_0$,
$r<b_0$, and $V_1(r)=0$, $r>b_0$) the binding-free energy $U(T)=-E_0$ takes
the form [20],
$$
U(T)=U_0f(T/T^*)
\eqno(2.7)
$$
where $T^*$ is an important characteristic temperature defined by
$$
T^*=\sqrt{\tilde\ep_1U_o}\;b_0\;.
\eqno(2.8)
$$
When $T\ll T^*$, the well depth  is effectively infinite, and we find
the usual particle-in-a-box result,
$$
U(T)\approx U_0-c_1{T^2\over
2\tilde\ep_1b_0^2}
\eqno(2.9)
$$
where $c_1$ is a constant related to the first zero of the Bessel function
$J_0(x)$ which solves Eq.~(2.4) in this limit. The localization length is then
$$
\ell_\perp(T)\approx
b_0[1+\CO(1/\kappa b_o)]\;,
\eqno(2.10)
$$
where $\kappa^{-1}\approx T/\sqrt{2\tilde\ep_1U(T)} \ll b_0$ is the distance
the ``particle'' penetrates into the classically forbidden region.
The low-temperature correction in Eq. (2.9) represents the entropy lost each
time a wandering flux line is reflected off the confining walls of the binding
potential [22]. At the crossover temperature $T^*$, this ``zero point energy''
of confinement becomes comparable to the well depth.

When $T\gg T^*$, the flux line is only weakly bound,
although a strictly localized ground state {\it always} exists
in this effectively two-dimensional problem [21].
In the limit, one finds [15,20]
$$
U(T)\approx {1\over 2}\;U_0\left({T\over T^*}\right)^2
e^{-2(T/T^*)^2}\;,
\eqno(2.11)
$$
and a localization length
$\ell_\perp(T)\approx\kappa^{-1}=T/\sqrt{2\tilde\ep_1U(T)}$, so that
$$
\ell_\perp(T)\approx b_0e^{(T/T^*)^2}\;.
\eqno(2.12)
$$
The flux line now ``diffuses'' within a confining tube with
radius of order $\ell_\perp
(T)$ as it crosses the sample. The length along $\hat{\bf z}$ required to
``diffuse'' across this tube is $\ell_z\approx\ell_\perp^2/(T/\tilde\ep_1)$,
i.e.,
$$
\ell_z=(b_0^2\tilde\ep_1/T)
e^{2(T/T^*)^2}\;.
\eqno(2.13)
$$

These results imply a strong thermal renormalization of the critical
current $J_c(T)$ [15]. $J_c(T)$ is the current necessary to produce
a Lorentz force $f_c=J_c\phi_0/c$ so strong that thermal activation
is unnecessary to tear a flux line away from its columnar pin. At low
temperatures, one would expect that $f_c\approx U_0/b_0$, i.e.,
$$
J_c\approx cU_0/\phi_0b_0\;.
\eqno(2.14)
$$
Here we assume that the confining potential $V_1(r)$ does not really jump
abruptly at $r=b_0$, but instead rises smoothly to zero in a distance of
order $b_0$. Results such as (2.11--2.13) are in any case
independent of the precise
form of the microscopic potential [21].
To account for line wandering, we should replace $U_0$ by $U(T)$ and
$b_0$ by $\ell_\perp(T)$. For $T\gg T^*$, this leads to
$$
J_c(T)\approx J_c(0)e^{-3(T/T^*)^2}\;.
\eqno(2.15)
$$

\onelnskp

\noindent
2.3. ONE FLUX LINE AND TWO COLUMNAR PINS

\onelnskp

\noindent
The localization length discussed above
is like the Bohr radius of an isolated ``atom'' consisting
of one columnar pin and one vortex line. Now consider a vortex line which is
able to hop between {\it two} nearby identical columnar pinning sites at
$\R_1$ and $\R_2$ (analogous to an $H_2^+$ molecule) as it traverses the
sample---see Fig.~5. For now, we ignore the additional dashed flux line.
The binding potential $V_1(r)$ in Eq.~(2.4) in this case should be replaced by
$$
V_2(\r)=V_1(\r-\R_1)+
V_1(\r-\R_2)\;,
\eqno(2.16)
$$
where identical wells of depth $U_0$ and radius $b_0$ separated
by $d$ are assumed. As in a
quantum double well, the dominant configurations are those in which the
particle delocalizes further by ``tunneling'' from one well to another. In the
subspace spanned by the isolated pin ground state wave functions $\psi_0(\r-
\R_1)$ and $\psi_0 (\r-\R_2)$, the effective Hamiltonian
takes the form
$$
\CH_2=\left(\matrix{E_0\;,&t\cr\;\;t\;,&E_0\cr}\right)\;,
\eqno(2.17)
$$
where $E_0$ is the ground state energy determined in the previous subsection,
and $t>0$ is a tunneling matrix element. The two lowest energy eigenvalues for
the double well are then
$$
E_\pm=E_0\pm t\;.
\eqno(2.18)
$$
The partition function in this approximation is $Z_{\rm tot}=
\sum_{i,j=1}^2\la i|\,e^{-\CH L/T}\,|j\ra$, where $|1\ra$ and $|2\ra$
are states localized on pin~1 or 2.

\topinsert
\centerline{\psboxto(0pt;5truein){fig5.eps}}
\onelnskp
\noindent
Figure 5. Two columnar pins with a single vortex (solid line) tunneling
between them. The dashed line shows a second vortex which requires the
introduction of a simple Hubbard model.
\onelnskp
\endinsert

To determine $t$, we can proceed variationally and minimize
$$
E(\psi)\equiv
{\int d^2r\left[
{T^2\over 2\tilde\epsilon_1}
|\nabla \psi(\r)|^2+V_2(\r)|\psi(\r)|^2\right]
\over
\int d^2r|\psi(\r)|^2}\;,
\eqno(2.19)
$$
with the trial function
$$
\psi(\r)=\alpha\psi_0
(\r-\R_1)+\beta
\psi_0(\r-\R_2)\;,
\eqno(2.20)
$$
where $\psi_0(\r)$ is the ground state for an isolated well.
We assume widely separated wells, i.e., $d\gg\ell_\perp(T)$.
The minimum occurs for the symmetric case $\alpha=\beta$ and has energy
$E=E_0-t$, with
$$
t\approx {\rm const.} \times
{U(T)\over\sqrt{E_k/T}}\;
e^{-E_k/T}\;,
\eqno(2.21)
$$
where $U(T)=-E_0(T)>0$, and
$$
E_k=\sqrt{2\tilde\epsilon_1 U(T)}\;d
\eqno(2.22)
$$
is similar to a WKB tunneling exponent.

The flux line has now delocalized a distance $\sim d$ in the transverse
direction. Delocalization proceeds via wandering in a
tube of radius $\ell_\perp(T)$,
and occasional tunneling across to a neighboring tube, as
indicated in Fig.~5. When $b_0\ll \ell_\perp\ll d$ the
spacing between such tunneling events along the $z$ axis is
of order
$$
\ell_z(T)\approx\left[
{b_0^2\tilde\ep_1\over T}\;
e^{2(T/T^*)^2}\right]
e^{E_k/T}\;,
\eqno(2.23)
$$
where the prefactor (an inverse ``attempt
frequency'' in imaginary time) comes from the
isolated pin result Eq.~(2.13). Note the close analogy between Fig.~5 and the
configurations of a classical one-dimensional Ising model with exchange
constant
$J=E_k$ disrupted by kinks along the $z$ axis.

The flux line will not be delocalized much further by adding a third, more
distant columnar pin to the problem, because the new available state
is not in resonance with the double well ground state energy calculated
above. One isolated flux line will always be localized by a random
array of columnar pins [15].

\onelnskp

\noindent
2.4. A HUBBARD MODEL: TWO FLUX LINES AND TWO COLUMNAR PINS

\onelnskp

\noindent
Interactions are crucial for determining vortex configurations when
correlated pinning is present. In absence of a repulsive pair potential
all vortices would pile up at $T=0$ in a deep minimum of the random
pinning potential produced, for example, by an unusually dense region of
columnar pins. By allowing {\it two} vortices to wander simultaneously
between a pair of columnar pins, we can study interactions in a
particularly simple
context. This elementary model also illustrates why flux lines
behave like bosons in thick samples. A related treatment describes
the $H_2$ molecule in real quantum mechanics [23].

Figure 5 shows two vortices (solid and dashed lines)
hopping back and forth between two columnar
pins as we trace their trajectories along the $z$ axis. In the
absence of interactions, the probability distribution of each
vortex would be described by the symmetric double well ground state wave
function discussed in the previous section. The vortices would
find themselves on the same columnar pin approximately half the
time. Introducing a repulsive energy between vortices will lead
to {\it correlated} hopping as the fluxons exchange places.

To model this situation, we introduce a tight-binding Hamiltonian
similar to Eq.~(2.17) which operates on a set of four
normalized orthogonal basis states, $|12\rangle$, $|21\rangle$,
$|11\rangle$, and $|22\rangle$. The state $|12\rangle$ means that
the first fluxon line occupies pin~1 and the second occupies pin~2,
while $|21\rangle$ is the state where the vortices have
exchanged places. Both fluxons are localized on pin~1 or on pin~2
in the states $|11\rangle$ and $|22\rangle$, respectively.
The classical partition function which gives the probability
of making a transition from one of these four states
$|a\rangle$ to a final state $|b\rangle$ across a slab of
thickness $L$ is then
$$
\CZ(a,b;L)=
\langle b|\,e^{-\CH L/T}\,|a\rangle\;,
\eqno(2.24)
$$
where the tight-binding model is defined by the $4\times4$ matrix
Hamiltonian
$$
\CH=\bordermatrix{&|12\rangle&|21\rangle&|11\rangle&|22\rangle\cr
|12\rangle&2E_0&0&t&t\cr
|21\rangle&0&2E_0&t&t\cr
|11\rangle&t&t&2E_0+V_{\rm int}&0\cr
|22\rangle&t&t&0&2E_0+V_{\rm int}\cr}\;.
\eqno(2.25)
$$

The diagonal terms include the one vortex binding energy $E_0$ discussed
in Sec.~2.2., and a ``Hubbard repulsion'' term $V_{\rm int}$ which represents
the energy which arises when two vortices occupy the same columnar pin. A
reasonable estimate of $V_{\rm int}$ when $d\ll \lambda_{ab}$ is [15]
$$
V_{\rm int}\approx 2\epsilon_0\ln(d/\xi_{ab})\;,
\eqno(2.26)
$$
reflecting the energy cost of one doubly quantized vortex as opposed to two
singly quantized vortices separated by $d$. For $d\gg \lambda_{ab}$,
$V_{\rm int}\approx 2\eps_0\ln(\lambda_{ab}/\xi_{ab})$. The
off-diagonal terms in (2.25)
reflect transitions between the four basis states caused by
hops of a single vortex. The Hamiltonian breaks into symmetric and
antisymmetric subspaces,
$$
{\baselineskip=11pt
\CH=\bordermatrix{&|-\rangle&&|+\rangle&|11\rangle&|22\rangle\cr
|-\rangle&2E_0&\vdots&0&0&0\cr
&\multispan{5}{\dotfill}\cr
|+\rangle&0&\vdots&2E_0&\sqrt{2}t&\sqrt{2}t\cr
|11\rangle&0&\vdots&\sqrt{2}t&2E_0+V_{\rm int}&0\cr
|22\rangle&0&\vdots&\sqrt{2}t&0&2E_0+V_{\rm int}\cr}}
\eqno(2.27)
$$
when $|12\ra$ and $|21\ra$ are eliminated in favor of the symmetrized states
$$
|-\ra = {1\over\sqrt{2}}\,(|12\ra-|21\ra)\;,
\eqno(2.28{\rm a})
$$
and
$$
|+\ra = {1\over\sqrt{2}}\,(|12\ra+|21\ra)\;.
\eqno(2.28{\rm b})
$$

The one dimensional subspace of states antisymmetric under vortex
interchange would describe spinless fermions in conventional quantum
mechanics. The three dimensional symmetric subspace is appropriate to
boson quantum mechanics. As $L\to\infty$, partition functions such as
(2.24) will be dominated by the smallest eigenvalue of $\CH$. The four
eigenvalues of (2.27) are easily found to be
$$
\eqalignno{
\lambda^*\,&= 2E_0\cr
\lambda_0\,&= 2E_0 + V_\rint\cr
\lambda_+\,&= 2E_0 + \half V_\rint +\half \sqrt{V_\rint^2+16t^2}\cr
\lambda_-\,&= 2E_0 + \half V_\rint -\half \sqrt{V_\rint^2+16t^2}\;.&(2.29)\cr}
$$
The eigenvalue $\lambda^*$ belongs to the fermion subspace, and is
unaffected by interactions. The remaining
eigenvalues are bosonic. As shown in Fig.~6, the lowest eigenvalue
is always $\lambda_-$, in the boson subspace. It differs from the fermion
eigenvalue $2E_0$ by an amount proportional to $t^2/V_{\rm int}$ as
$V_{\rm int}\to\infty$.

\topinsert
\centerline{\psboxto(0pt;5truein){fig6.eps}}
\onelnskp
\noindent
Figure 6. Energy eigenvalues for the four state Hubbard model. Solid lines
represent the symmetric ``boson'' subspace, while the dashed line
corresponds to the antisymmetric ``fermion'' excitation.
\onelnskp
\endinsert

The boson character of the ground state is a general feature of Hamiltonians
which are symmetric under particle interchange [16]. It arises in the present
context because repeated hopping between the columnar pins leads to a
probability distribution symmetric in the two vortex coordinates for
sufficiently thick samples. The pair of flux lines in Fig.~6 will behave
like bosons for sample thicknesses $L>>\ell_z$, where $\ell_z(T)$ is the
spacing between these tunneling events.

The eigenvector of the ground state is
$$
|0\ra={\rm const.}\times\left[{1\over\sqrt2}\,|+\ra \;+\;
{1\over2}\left(\sqrt{1+\left(V_\rint\over 4t\right)^2}-{V_\rint\over 4t}\right)
(|11\ra+|22\ra)\right]\;.
\eqno(2.30)
$$
Note that the probability of double occupancy of a columnar pin vanishes like
$(t/V_{\rm int})^2$ as $V_{\rm int}\to\infty$.

Although the smallest eigenvalue of the transfer matrix $V$ dominates
the partition function as $L\to\infty$, the excited states in Fig.~6 are
important for correlation functions connecting different values of $z$ and
when $L\lot\ell_z$. Only states in the {\it bosonic} subspace contribute even
in this case, however. To understand this, recall that Eq.~(2.24) must be
summed over states $|a\ra$ and $|b\ra$ describing the entry and exit
points of the vortices. The total partition function is thus
$$
Z_{\rm tot}=\la\alpha|\,e^{-\CH L/T}\,|\alpha\ra\;,
\eqno(2.31)
$$
where $|\alpha\ra$ is some linear combination of $|11\ra$, $|22\ra$,
$|12\ra$ and $|21\ra$. Because the flux lines are indistinguishable,
the initial state must take the form
$$
|\alpha\ra=a|11\ra+b|22\ra+c(|12\ra+|21\ra)\;,
\eqno(2.32)
$$
where $a$, $b$, and $c$ are constants. We allow for $a\ne b$ because the
columnar pins could in fact have slightly different binding energies both
at the surface or in the interior, as in the Anderson model of localization
[15]. Even in this case, the initial state $|\alpha\ra$ lies completely within
the {\it boson} subspace of Eq.~(2.27). Thus the fermion eigenstate never
contributes to the statistical mechanics. Similar arguments show that only
boson excitations are relevant to arbitrarily large assemblies of interacting
flux lines [17,20].

\firstskip

\noindent
{\bf 3. Flux Melting and the Quantum Harmonic Oscillator}

\onelnskp

\noindent
Consider one representative fluxon in the confining potential ``cage''
provided by its surrounding vortices in a triangular lattice.  The
partition function for a fixed entry point $\0$ and exit point $\r_\perp$
in a sample of thickness $L$ is
$$
\eqalignno{
\CZ_1(\r_\perp,\0;L)&\,=
\int_{\r(0)=\0}^{\r(L)=\r_\perp}
\CD\r(z)\exp\left\{-{1\over T}\int_0^L
\left[\half\tel\left({d\r(z)\over dz}\right)^2+V_1[\r(z)]\right]dz\right\},\cr
&&(3.1)\cr}
$$
where $V_1[\r]$ is now a one-body potential chosen to mimic the interactions
in Eq.~(2.1a). We assume clean samples and high temperatures so that
both correlated and point disorder can be neglected.

Three important field regimes for fluctuations in vortex crystals are easily
extracted from this simplified model. Following the approach in the previous
section, we rewrite this imaginary time path integral
as a quantum mechanical matrix element,
$$
\CZ(\r_\perp,\0;L)=\langle\r_\perp| e^
{-L\CH/T}|\0\rangle\;,
\eqno(3.2)
$$
where $|\0\rangle$ is an initial state localized at $\0$, $\langle\r_\perp|$ is
a final state localized at $\r_\perp$, and the ``Hamiltonian'' $\CH$ is
the operator which appears in Eq.~(2.4),
$$
\CH=-{T^2\over 2\tel}\;
\nabla_\perp^2+V_1(\r)\;.
\eqno(3.3)
$$
Recall that the probability of finding the flux line at transverse position
$\r$ within the
crystal is  $\psi_0^2(\r)$, where $\psi_0(\r)$ is the
normalized ground state eigenfunction of (3.3).

When $B\gg B_{ c1}\equiv \phi_0/\lambda_{ab}^2$, the pair potential
$V(r_{ij})=2\epsilon_0K_0(r_{ij}/\lambda_{ab})$
is logarithmic, $K_0(x)\approx\ln x$, and we expand $V_1(\r_\perp)$
about its minimum at $\r_\perp=0$ to find
$$
\left[-{T^2\over 2\tel}\;
\nabla_\perp ^2+{1\over 2}\;
kr_\perp^2\right]
\psi_0=E_0\psi_0
\eqno(3.4)
$$
where (neglecting logarithmic corrections to $\ep_0$
and constants of order unity)
$$
\eqalignno{
k&\left.\approx {d^2V\over dr^2}\right|_{r=a_0}\cr
&\approx{\eps_0\over a_0^2}\;,
&(3.5)\cr}
$$
and $a_0$ is the mean vortex spacing. Equation (3.4) is the Schroedinger
equation for a two-dimensional quantum oscillator, with $\hbar\rightarrow T$
and mass
$m\rightarrow\tel$. The ground state wave function is
$$
\psi_0(r_\perp)=
{1\over\sqrt{2\pi}\;r_*}\;
e^{-r^2/4r_*^2}
\eqno(3.6)
$$
with spatial extent
$$
r_*=\left({T^2a^2_0\over
\eps_0\tel}\right)^{1/4}\;.
\eqno(3.7)
$$
Melting occurs when $r_*=c_La_0$, where $c_L$ is the Lindemann constant, so the
melting temperature is
$$
T_m=c_L^2\sqrt{\eps_0\tel}\;a_0\;,
\qquad\qquad\qquad  (B_{ c1}< B\;\lot\;B_x)
\eqno(3.8)
$$
in agreement with other estimates [24]. Vortices in the crystalline phase will
travel across their confining tube of radius $r_\perp^*$ in a ``time''  along
the $z$ axis of order $\ell_z^0$, where [17]
$$
\eqalignno{
\ell_z^0&\approx r_*^2/(T/\tel)\cr
&\approx \sqrt{{\tel\over\eps_0}}\;
a_0\;.
&(3.9)\cr}
$$

A new high field regime arises when $\ell_z^0\;\lot\;d_0$, where $d_0$ is the
average spacing of the copper-oxide planes, i.e., for $B\;\got\; B_x$,
with decoupling field
$$
B_x\approx{\tel\over \eps_0}\;
{\phi_0\over d_0^2}\;,
\eqno(3.10)
$$
again in agreement with earlier work [8, 24].
Above this field, the planes are approximately decoupled, and $T_m$ may be
estimated from the theory of two-dimensional dislocation mediated melting
[8, 24]
$$
T_m\approx {\eps_0d_0\over
8\pi\sqrt 3}\;,
\qquad\qquad\qquad (B\;\got\;B_x)\;.
\eqno(3.11)
$$

The estimate (3.8) also breaks down at low fields $B\;\lot\;B_{c1}$ where the
logarithmic interaction potential must be replaced by an exponential repulsion.
The two-dimensional harmonic oscillator model again applies, with the
replacement
$$
k\rightarrow {\eps_0\over\lambda_{ab}^2}\;
e^{-a_0/\lambda_{ab}}.
\eqno(3.12)
$$
The transverse wandering distance is now
$$
r_*\approx\left(
{T^2\lambda_{ab}^2\over\eps_0\tel}\right)
^{1/4} e^{a_0/4\lambda_{ab}}.
\eqno(3.13)
$$
and takes place over a longitudinal distance
$$
\ell_z^0=\sqrt{{\tel\over\eps_0}}\;
\lambda_{ab} e^{a_0/2\lambda_{ab}}.
\eqno(3.14)
$$
The low field melting temperature becomes
$$
T_m\approx c_L^2\sqrt{\eps_0\tel}
\;{a_0^2\over\lambda_{ab}}
e^{-a_0/2\lambda_{ab}}\qquad\qquad \qquad
(B\;\lot\;B_{ c1})\;,
\eqno(3.15)
$$
consistent with earlier predictions [17,8]. Although we have retained the
distinction between $\tel$ and $\eps_0$ in these formulas, note that
$\tel\approx\eps_0$ in this regime [8].

\topinsert

\noindent
\parindent=0pt
Table 1.  Estimates for the flux lattice melting temperature determined for
the three regimes discussed in the text

\onelnskp

\tabskip=1em plus 2em minus .5em
\halign to \hsize{\hfil#\hfil&&\hfil#\hfil\cr
\noalign{\hrule}
\noalign{\vskip1pt}
\noalign{\hrule}
\noalign{\vskip6pt}
Regime&$T_m(B)$\cr
\noalign{\vskip4pt}
\noalign{\hrule}
\noalign{\vskip6pt}
$B_x\;\lot\;B$&$\eps_0d_0/8\pi\sqrt3$\cr
&&$B_x\approx{\tel\over\eps_0}\,{\phi_0\over d_0^2}$\cr
$B_{c1}\lot B\;\lot\;B_x$&$c_L^2\sqrt{\eps_0\tel}(\phi_0/B)^{1/2}$\cr
&&$B_{c1}\approx \phi_0/\lambda^2_{ab}$\cr
$B\;\lot\;B_{c1}$&$c_L^2\sqrt{\eps_0\tel}\lambda_{ab}\left({B_{c1}\over
B}\right)e^{-{1\over 2}(B_{c1}/B)^{1/2}}$\cr
\noalign{\vskip6pt}
\noalign{\hrule}
\noalign{\vskip1pt}
\noalign{\hrule}
\cr}
\onelnskp

\endinsert

The predictions (3.8), (3.11) and (3.15) are combined to give the reentrant
phase diagram for melting shown in Fig.~1.
Analytic estimates and boundaries for
melting in the various regimes are summarized in Table~1.

\firstskip

\noindent
{\bf 4. Vortex Entanglement in the Liquid Phase}

\onelnskp

\noindent
Above the melting line in Fig.~1, weak point disorder due to oxygen vacancies
can usually be neglected [7] and we must consider a liquid of wandering,
essentially unconfined lines. To estimate the degree of entanglement, we
consider a {\it single} flux line $\br(z)$ and determine how far it wanders
perpendicular to the $z$ axis as it traverses the sample. The relevant path
integral is
$$
\la|\br(z)-\br(0)|^2\ra=
{\int\CD\br(s)|\br(z)-\br(0)|^2\exp
\left[-{\tilde\ep_1\over 2T}\int\nolimits_0^L
\left(d\br\over ds\right)^2\,ds\right]\over
\int\CD\br(s)\exp
\left[-{\tilde\ep_1\over 2T}\int\nolimits_0^L
\left(d\br\over ds\right)^2\,ds\right]}\;,
$$
which, when discretized as in Appendix~A, yields
$$
\la|\br(z)-\br(0)^2\ra=
{2T\over\tilde\ep_1}\,|z|\;,
\eqno(4.1)
$$
which shows that the vortex ``diffuses'' as a function of the timelike
variable $z$,
$$
\la|\br(z)-\br(0)|^2\ra^{1/2}=(2Dz)^{1/2}\;,
\eqno(4.2)
$$
with diffusion constant
$$
D={T\over\tilde\ep_1}={M_z\over M_\perp}\,{4\pi T\over\phi_0H_{c1}}\;.
\eqno(4.3)
$$
At $T=77$K, we take $H_{c1}\approx 10^2$~G and $M_z/M_\perp\approx10^2$ and
find $D=10^{-6}$~cm, so that vortex lines wander a distance of order 1~$\mu$m
while traversing a sample of thickness 0.01~cm.

Close encounters between neighboring vortex lines will thus occur quite
frequently in fields of order 1~T or more, where vortices are separated
by distances of order 500~\AA\ or less. The ``entanglement length'' $\ell_z$
is defined as the distance along the $z$ axis such that
$\la|\r(\ell_z)-\r(0)|^2\ra=a_0^2=B/\phi_0$, i.e.\ [17]
$$
\ell_z={a_0^2\over 2D}={\tilde\ep_1 a_0^2\over2T}\;.
\eqno(4.4)
$$
Collisions and entanglement of vortex lines wil be important for flux liquids
whenever
$$
L>\ell_z\;,
\eqno(4.5)
$$
i.e., for $B>B_{x1}\approx(M_\perp/M_z)(\phi_0\ep_0/LT)$.

The discrete flux filaments which comprise the flux liquid first form when
a superconductor is cooled through the mean field transition line
$B^0_{c2}(T)$.
These fluxons are ``phantom vortices'' for $B\lot B^0_{c2}(T)$ in the sense
that the barriers to line crossing are expected to be negligible. The
barriers will increase, however, as the flux liquid is cooled toward the
fluctuation induced melting temperature of the Abrikosov flux crystal.
Entanglement can then cause the flux liquid to become very viscous [17,26].

The consequences for transport of vortices in the vicinity of a few strong
pinning centers such as twin boundaries can be quite striking. Assume that
a supercurrent $\j_s$ flows along $\hat{\bf y}$, $\j_s=j_s\hat {\bf y}$,
and that the magnetic field, as usual, is parallel to $\hat{\bf z}$. Vortices
will then be subjected to a constant driving Lorentz force,
$$
\f_L={1\over c}\,n_0\phi_0\hat z\times\j_s \;,
\eqno(4.6)
$$
where $n_0\approx a_0^{-2}$ is the vortex density.
The equation of motion which describes a $z$-independent flux liquid velocity
field in the vicinity of, say, a twin boundary in the $xz$ plane, is then [26]
$$
-\gamma\vv +\eta\nabla_\perp^2\vv+\f_L=0\;.
\eqno(4.7)
$$
Eq.~(4.7) is a hydrodynamic description of flux flow, valid on large scales
compared to the intervortex spacing. The Bardeen-Stephen parameter $\gamma$
is a ``friction'' coefficient which represents the resistance encountered
to vortex motion by the normal electrons in the core. The combination of the
drag and viscous terms in Eq.~(4.7) introduces an important new length scale
into the problem,
$$
\delta=\sqrt{\eta/\gamma}\;.
\eqno(4.8)
$$
The length $\delta$ is the scale over which the velocity rises to its bulk
value from the center of the twin boundary where it is small or vanishes
entirely.

The flux line viscosity has been estimated, e.g., by Cates [27], who finds
a remarkably simple formula for $\delta$
$$
\delta\approx a_0e^{U_\times/T}
\eqno(4.9)
$$
where $U_\times$ is the barrier to line crossing. We estimate this barrier as
shown in Fig.~7, following the treatment of Obukhov and Rubinstein [28].
The projections of two vortices (displaced initially by $\sim a_0$
perpendicular to the plane of the figure) intersect near the center of the
sample and approach the intervortex spacing at $z=\pm L/2$. Crossing
is very difficult at low angles $\theta$, because of the extra energy
associated with a doubly quantized filament formed near the crossing region
[17]. Crossing is easier at high angles, however. In an isotropic
superconductor, the crossing energy goes to zero when $2\theta=90$\deg,
for example [29]. A simple rescaling argument shows that this critical angle
for zero crossing energy becomes
$$
\theta_c\approx \tan^{-1}(\sqrt{M_z/M_\perp})
\eqno(4.10)
$$
for anisotropic materials in the symmetrical situation shown in Fig.~7.
One way to exchange the lines shown in Fig.~7a is to pass to the intermediate
configuration shown schematically in Fig.~7b, where the vortices bend and
acquire extra length so they are inclined at the critical angle. The
filaments can then pass through each other remaining crossed,
relaxing finally to the configuration shown in Fig.~7c.

\topinsert
\centerline{\psboxto(\hsize;0pt){fig7.eps}}
\onelnskp
\noindent
Figure 7. A mechanism for vortex line crossing. Vortices in the initial
(a) and final (c) configurations are displaced out of the plane of the
figure by a distance $\sim a_0$. An intermediate saddle point configuration
connecting these states is shown in (b). Here $\gamma=(M_\perp/M_z)^{1/2}$.
\onelnskp
\endinsert

The line energy of a wandering vortex can be parameterized in terms of its
arc length $s$ as
$$
\int_0^L \ep_1(\theta){ds\over dz}\,dz\;,
\eqno(4.11)
$$
where $\theta(z)$ is the local angle of the inclination relative to $z$ and
[30]
$$
\ep_1(\theta)=\ep_1\sqrt{\cos^2\theta+{M_\perp\over M_z}\,\sin^2\theta}
\eqno(4.12)
$$
with $\ep_1=\ep_0\ln(\lambda_{ab}/\xi_{ab})$. For the configuration in
Fig.~7a, this formula leads immediately to an energy
$$
E_0=2\ep_1L[1+\CO(a_0^2/L^2)]\;.
\eqno(4.13)
$$
The saddle point energy shown in Fig.~7b, on the other hand, has energy
$$
\eqalignno{
E_0'\,&=2\ep_1\left(L-\sqrt{M_\perp\over M_z} a_0\right)+
2\ep_1
\sqrt{\cos^2\theta_c+{M_\perp\over M_z}\,\sin^2\theta_c}
\sqrt{a_0^2+(\cot^2\theta_c)a_0^2}\cr
&&(4.14)\cr}
$$
which simplifies using Eq.~(4.10) to
$$
E'_0=2\ep_1 L+2(\sqrt2-1)\,\sqrt{M_\perp\over M_z} a_0\ep_1\;.
\eqno(4.15)
$$
The change in the interaction energy between the lines due to the sharp bends
and nonzero tilt will add a correction to the coefficient of the second
term, which we neglect. Note also that the terms neglected in (4.13) are
higher order in $a_0/L$ than the terms kept in (4.15). The crossing energy
$U_\times\approx E'_0-E_0$ is thus approximately [28]
$$
U_\times\approx 2(\sqrt2-1)\sqrt{M_\perp\over M_z} a_0\ep_1\;.
\eqno(4.16)
$$

We see that the crossing energy tends to zero as $M_\perp/M_z\to 0$, and has
essentially the same functional form as the intermediate field melting
temperature displayed in Eq.~(3.8). The key parameter which determines the
viscous length scale (4.9) at the melting temperature is then
$$
{U_\times\over T_m}\approx{c_\times\over c_L^2}\;,
\eqno(4.17)
$$
where $c_\times$ is a dimensionless constant, $c_\times=(2\sqrt2-1)\ln
(\lambda_{ab}/\xi_{ab})$, for the simple model discussed above.
This dimensionless ratio should be independent of field strength in the range
$B_{c1}\ll B\ll B_x$. If $U_\times/T_m\lot 1$, the crossing barriers associated
with entanglement will not interfere with crystallization into an Abrikosov
flux lattice. (This should {\it always} be the case for $B\gg B_x$.) Note,
however, that it is the Lindemann ratio squared which enters the denominator
of (4.17). Assume for concreteness that $c_\times=0.75$.
If $c_L=0.3$, then $U_\times/T_m\approx8$ and $\delta=a_0e^{U_\times/T_m}
\approx4\times10^3a_0$. If $c_L=.15$,
as indicated in the most recent experiments on untwinned YBCO [3], then
$U_\times/T_m=33$, and $\delta\approx3\times10^{14}a_0$! If
$U_\times/T_m$ is really this large, the kinetic barriers associated with
entanglement will {\it preclude} crystallization on experimental time scales
in samples thick enough or fields high enough to allow multiple entanglements
of the vortex lines. The flux liquid will instead form a ``polymeric glass''
phase upon cooling [17,28]. It is interesting to note for comparison purposes
that $U_\times/T_m\approx75$ in a {\it real} polymer like polyethyene. More
accurate estimates of the numerator of (4.17) would, of course, be highly
desirable.

Some evidence for the ``polymer glass'' scenario already exists. Recent
experiments by Safar \etal\ find that the first order melting transition
in untwinned YBCO goes away in sufficiently high magnetic fields,
$B\got 10$T. One possibility is that point disorder somehow becomes
more important and causes a ``vortex glass'' transition [8] at high
fields [31]. An alternative explanation, however, is that the flux liquid
simply becomes more entangled and viscous as its density increases with
increasing field, leading eventually to undercooling and a polymer glass
transition.  Recent simulations of a lattice superconductor have revealed
large crossing barriers and are never able to recover the crystalline phase
upon cooling
once the flux lattice has melted [32], in agreement with this picture.
In real experiments, the hysterisis loops in the resistivity associated with
first order freezing would slowly go away as crystallization became more
difficult for a fixed cooling rate. The low temperature dynamics of this
polymer glass in the presence of point disorder should be similar to that
predicted by the collective pinning theory [33], with a {\it polymeric}
shear modulus replacing the usual crystalline
flux lattice elastic constant $c_{66}$.

\firstskip

\noindent
{\bf Acknowledgements}

\onelnskp

\noindent
This work was supported by the National Science Foundation, through Grant
No.\ DMR~91-15491 and through the Harvard Materials Research Laboratory.
Much of this work is the result of stimulating collaboration with V.M. Vinokur.
See Ref.~15 for a more complete treatment of
correlated pinning. Discussions with D.~Bishop,
R.~Budhani, L.~Civale, G.~Crabtree, D.S.~Fisher, P.L.~Gammel, T.~Hwa,
P.~Le~Doussal and M.C.~Marchetti, are also gratefully acknowledged.

\firstskip

\def\v0{{\vec 0}}
\def\A{{\rm A}}
\def\vpr{^{\,\prime}}

\noindent
{\bf Appendix A: Transfer Matrix Representation of the Partition Function}

\onelnskp

\noindent
We review here how path integrals like those represented in Eq.~(2.3) can be
rewritten in terms of a transfer matrix, which is the exponential of the
Schroedinger operator which appears in elementary quantum mechanics [16].

We first consider
$$
\eqalignno{
Z(\r_\perp,\v0;\ell)\,&=
\int_{\r(0)=\v0}^{\r(\ell)=\r_\perp} \CD\r(z)
\exp\left[-{\tilde\ep_1\over 2T}\,\int_0^\ell
\left({d\r\over dz}\right)^2\,dz
-{1\over T}\int_0^\ell V_1[r(z)]\,dz\right]\cr
&&(\A1)\cr}
$$
and discretize this path integral as indicated in Fig.~8, where the planes
of constant $z$ are separated by a small parameter $\delta$. This is
precisely the situation which arises in the high $T_c$ superconductors
with field parallel to the $c$ axis, provided $\delta$ represents the mean
spacing between CuO$_2$ planes. The discretized path integral reads
$$
\eqalignno{
Z(\r_\perp,\v0,\ell)\approx
\left(\prod_{j=2}^N{\tilde\ep_1\over2\pi T\delta}\int d^2r_j\right)
\exp\left[{-\tilde\ep_1\over2T\delta}\,\sum_{j=2}^{N+1}(\r_j-\r_{j-1})^2-
{\delta\over T}\,\sum_{j=1}^{N+1} V_1(\r_j)\right]\;,\cr
&&(\A2)\cr}
$$
where it is understood that $\r_1=\v0$, $\r_{N+1}=\r_\perp$, and the
normalization of the integrals is chosen so that $Z(r_\perp,0;\ell)=1$
when $V_1(\r)=0$.

\topinsert
\centerline{\psboxto(\hsize;0pt){fig8.eps}}
\onelnskp
\noindent
Figure 8. Discretized path integral representation of vortex line passing
through $N+1$ CuO$_2$ planes with average spacing $\delta=\ell/N$.
\onelnskp
\endinsert

As usual in statistical mechanics, the effect of adding one copper oxide plane
to the system can be represented in terms of a transfer matrix
$$
Z(\r,\v0;\ell+\delta)=
\int d^2 r' T(\r,\r^{\,\prime})\,Z(\r\vpr,\v0;\ell)
\eqno(\A3)
$$
where (neglecting small edge effects at the top and bottom of the sample)
$$
T(r,\r\vpr)={\tilde\ep_1\over2\pi T\delta}\,
\exp\left\{{-\tilde\ep_1\over2T\delta}\,|\r-\r\vpr|^2-
{\delta\over 2T}\,[V_1(\r)+V_1(\r\vpr)]\right\}\;.
\eqno(\A4)
$$
Ryu \etal\ have studied the spectrum of $T(\r,\r\vpr)$ for finite
$\delta$ [34]. Here, we shall instead consider the limit of small $\delta$.
We can then expand the potential term and derive a differential equation
for $Z$,
$$
T\partial_\ell Z(\r,\v0;\ell)=
\left[{T^2\over2\tilde\ep_1}\,\nabla^2_\perp-V_1(\r)\right]\,
Z(\r,0;\ell)\;.
\eqno(\A5)
$$
A formal expression for the partition function results from integrating
Eq.~(A5) across a sample thickness $L$ with initial state $|0\ra$ and
final state $|\r_\perp\ra$,
$$
Z(\r_\perp,0;\ell)=\la\r_\perp|\,e^{-\CH L/T}\,|\v0\ra\;,
\eqno(\A6)
$$
where $\CH={-T^2\over2\tilde\ep_1}\,\nabla_\perp^2+V(\r)$.
The partition function $Z(\v0,\r_\perp;L)$ has been defined to be the
ratio of the path integrals which appear in Eq.~(2.3). Upon inserting
a complete set of normalized eigenfunctions $\{\psi_n(\r)\}$ of $\CH$
with energies $\{E_n\}$ into (A6), we obtain
$$
\eqalignno{
e^{U(T)L/T}\,&=Z(\r_\perp,0;L)&\cr
&=\sum_n\psi(\r_\perp)\psi(\v0)e^{-E_nL/T}&(\A7)\cr}
$$
which leads when $L\to\infty$ to the result
$$
U(T)=-E_0(T)
\eqno(\A8)
$$
quoted in the text.

\firstskip

\noindent
{\bf Appendix B: Vortex Probability Distributions}

\noindent
We show here that the probability of finding an individual vortex line at
height
$z$ with position $\r_\perp$ in an arbitrary binding potential $V_1(\r_\perp)$
is related to the ground state wave function of the corresponding
Schroedinger equation. The probability distribution at a free
surface is proportional to the wave function itself, while the probability
far from the surface is proportional to the wave function squared.

Consider first a fluxon which starts at the origin $\0$ and wanders across a
sample of thickness $L$ to position $\r_\perp$. As discussed in Appendix~A,
the partition function associated with this constrained path integral may
be written as a quantum mechanical matrix element
$$
\eqalignno{
\CZ(\r_\perp,\0;L)&=
\int_{\r(0)=\0}^{\r(L)=\r_\perp}
\CD r(z)\exp
\left[
-{\tilde\ep_1\over 2T}
\int_0^L\left|{d\r\over dz}\right|^2dz-
{1\over T}\int_0^L V_1[\r(z)]dz\right]\cr
&\equiv \langle \r_\perp| e^{-L\CH/T}|0\rangle
&({\rm B1})\cr}
$$
where $|\0\rangle$ is an initial state localized at $\0$ while
$\langle\r_\perp|$ is a final state localized at $\r_\perp$. The
``Hamiltonian'' $\CH$ appearing in (B1) is the Schroedinger operator,
$$
\CH=-{T^2\over 2\tilde\ep_1}\;
\nabla_\perp^2+ V_1(\r)\;.
\eqno({\rm B2})
$$
The probability distribution $\CP(\r_\perp)$ for the vortex tip
position at the upper surface is then
$$
\CP(\r_\perp)=\CZ(\r_\perp,\0;L)\Bigg/\int d^2r_\perp
\CZ(\r_\perp,\0; L)\;.
\eqno({\rm B3})
$$
Upon inserting a complete set of (real) energy eigenstates $|n\rangle$ with
eigenvalues $E_n$ into Eq.~(B1), we have
$$
\CP(\r_\perp)=
{\sum_n
\psi_n(\0)
\psi_n(\r_\perp)
e^{-E_nL/T}\over
\sum_n\psi_n(0)\int d^2r_\perp\psi_n(\r_\perp)
e^{-E_nL/T}}\;.
\eqno({\rm B4})
$$
In the limit $L\rightarrow \infty$ the ground state dominates, the probability
$\CP(\r_\perp
)$ becomes
$$
\CP(\r_\perp)\approx
{\psi_0(\r_\perp)\over\int
d^2r_\perp\psi_0(\r_\perp)}
\left[1+\CO\left(
e^{-(E_1-E_0)L/T}\right)
\right]\;,
\eqno({\rm B5})
$$
where $E_1$ is the energy of the first excited state.
Because the ground wave function is nodeless [21], $\CP(r_\perp)$ is always
positive and well defined.

Consider now a more general problem of a vortex which enters the sample at
$\r_i$, exits at $\r_f$, and passes through $\r$ at a height $z$ which is
far from the boundaries. The normalized probability distribution is now
$$
\tilde\CP(\r;L)=
\tilde\CZ(\r;L)/
\int d^2r\tilde\CZ(\r;L)
\eqno({\rm B6})
$$
where
$$
\tilde\CZ(\r;L)=
\int d^2r_i\int d^2r_f\CZ(\r_f,\r;L-z)
\CZ(\r,\r_i;z)
\eqno({\rm B7})
$$
and $Z(\r_2,\r_1;L)$ is given by Eq.~(B1).
Upon inserting complete sets of states as before, we find that
$$
\tilde\CP(\r;L)=
{\psi_0^2(r)\over
\int d^2r_\perp\psi_0^2(r)}
\left[1+\CO\left(e^{-L(E_1-E_0)/2T}
\right)\right]\;,
\eqno({B8})
$$
where the correction assumes $\r$ is at the midplane of the sample.

\firstskip

\centerline{\bf References}
\medskip

\item{1.} A.A. Abrikosov, {\it Zh.\ Eksperim. i Theor.\ Fiz. \bf 32}, 1442
(1957) [{\it Sov.\ Phys.\ JETP \bf 5}, 1174 (1957)].
\item{2.} M. Charalmabous \etal, {\it Phys.\ Rev.\ \bf B45}, 45 (1992).
\item{3.} W.K. Kwok, S. Fleshler, U. Welp, V.M. Vinokur, J. Downey,
G.W. Crabtree and M.M. Miller,
{\it Phys.\ Rev.\ Lett.\ \bf 69}, 3370 (1992).
\item{4.} D.E. Farrell, J.P. Rice and D.M. Ginsberg,
{\it Phys.\ Rev.\ Lett.\ \bf 67}, 1165 (1991).
\item{5.} H. Safar, P.L. Gammel, D.A. Huse, D.J. Bishop, J.P. Rice and
D.M. Ginsberg, {\it Phys.\ Rev.\ Lett.\ \bf 69}, 824 (1992).
\item{6.} E. Brezin, D.R. Nelson and A. Thiaville,
{\it Phys.\ Rev.\ \bf B31}, 7124 (1985).
\item{7.} D.R. Nelson and P. Le Doussal,
{\it Phys.\ Rev.\ \bf B42}, 10112 (1990).
\item{8.}  D.S. Fisher, M.P.A. Fisher and D.A. Huse,
{\it Phys.\ Rev.\ \bf B43}, 130 (1991).
\item{9.} R.C. Budhani, M. Suenaga and H.S. Liou,
{\it Phys.\ Rev.\ Lett.\ \bf 69}, 3816 (1992).
\item{10.} M. Konczykowski \etal, {\it Phys.\ Rev.\ \bf B44}, 7167 (1991).
\item{11.} L. Civale, A.D. Marwich, T.K. Worthington, M.A. Kirk, J.R. Thompson,
L. Krusin-Elbaum, Y. Sun, J.R. Clem and F. Holtzberg,
{\it Phys.\ Rev.\ Lett.\ \bf 67}, 648 (1991).
\item{12.} W. Gerhauser \etal, {\it Phys.\ Rev.\ Lett.\ \bf 68}, 879 (1992).
\item{13.} V. Hardy \etal, {\it Nucl.\ Instr.\ and Meth.\ \bf B54}, 472 (1991).
\item{14.} A.P. Malozemoff, T.K. Worthington, Y. Yeshurun and F. Holtzberg,
{\it Phys.\ Rev.\ \bf B38}, 7203 (1988).
\item{15.} D.R. Nelson and V.M. Vinokur,
{\it Phys.\ Rev.\ Lett.\ \bf 68}, 2392 (1992), and Harvard University preprint.
\item{16.} R.P. Feynman and A.R. Hibbs, {\it Quantum Mechanics and Path
Integrals} (McGraw-Hill, New York, 1965); R.P. Feynman, {\it Statistical
Mechanics} (Benjamin, Reading, MA, 1972).
\item{17.} D.R. Nelson, {\it Phys.\ Rev.\ Lett.\ \bf 60}, 1973 (1988);
D.R. Nelson and S. Seung, {\it Phys.\ Rev.\ \bf B39}, 9153 (1989).
\item{18.} S.F. Edwards, {\it Proc.\ Phys.\ Soc.\ \bf 85}, 613 (1965);
P.G. de Gennes, {\it Rep.\ Prog.\ Phys.\ \bf 32}, 187 (1969).
\item{19.} M.P.A. Fisher, P.B. Weichman, G. Grinstein and D.S. Fisher,
{\it Phys.\ Rev.\ \bf B40}, 546  (1989), and references therein.
\item{20.} D.R. Nelson, in {\it Phenomenology and Applications of High
Temperature Semiconductors}, edited by K. Bedell, M. Inui, D. Meltzer,
J.R. Schrieffer, and S. Doniach (Addison-Wesley, New York, 1991).
\item{21.} L.D. Landau and E.M. Lifshitz, {\it Quantum Mechanics},
2nd.\ Edition (Pergammon, New York, 1965).
\item{22.} See, e.g., D.R. Nelson, {\it J.\ Stat.\ Phys.\ \bf 57}, 511 (1989).
\item{23.} N.W. Ashcroft and N.D. Mermin, {\it Solid State Physics}
(Sanders College, Philadelphia, 1976), Chapter~32.
\item{24.} See, e.g., L.I. Glazman and A.E. Koshelev,
{\it Phys.\ Rev.\ \bf B43}, 2835 (1991).
\item{25.} D.S. Fisher, {\it Phys.\ Rev.\ \bf B22}, 1190 (1980).
\item{26.} M.C. Marchetti and D.R. Nelson, {\it Phys.\ Rev.\ \bf B42},
9938 (1990); {\it Physica \bf C174}, 40 (1991).
\item{27.} M. Cates, {\it Phys.\ Rev. \bf B45}, 12415 (1992).
\item{28.} S. Obukhov and M. Rubinstein,
{\it Phys.\ Rev.\ Lett.\ \bf 66}, 2279 (1991); see also
S. Obukhov and M. Rubinstein, {\it Phys.\ Rev.\ Lett.\ \bf 65}, 1279 (1990).
\item{29.} E.H. Brandt, J.R. Clem and D.G. Walmsley,
{\it J.\ Low Temp.\ Phys.\ \bf 37}, 43 (1979).
\item{30.} V.G. Kogan,
{\it Phys.\ Rev. \bf B24}, 1572 (1981).
\item{31.} H. Safar, P.L. Gammel, D.A. Huse, D.J. Bishop, W.C. Lee,
J. Giapintzakis and D.M. Ginsberg, AT\&T Laboratories preprint.
\item{32.} S. Teitel, private communication.
\item{33.} M.V. Feigel'man, V.B. Geshkenbein, A.I. Larkin and V.M. Vinokur,
{\it Phys.\ Rev.\ Lett.\ \bf 63}, 2303 (1989).
\item{34.} S. Ryu, A. Kapitulnik and S. Doniach, Stanford University preprint.

\autojoin
\bye